\documentclass{aa}  

\newcommand\dosingle[1]{#1}  \newcommand\dodouble[1]{ } 

\newcommand\nice[1]{#1}    \newcommand\subm[1]{}   
\subm{ \renewcommand\dosingle[1]{ }   \renewcommand\dodouble[1]{#1} }

\dodouble{ \documentclass[referee]{aa} }

\usepackage{graphicx}

\usepackage{amsfonts}
\usepackage{mathpi}

\usepackage{url}

\newcommand\prerefereechanges[1]{#1}  \newcommand\prerefereestart{  }  \newcommand\prerefereestop{ }

 \usepackage{hyperref}
\nice{ 
\providecommand{\eprint}[1]{\href{http://arxiv.org/abs/#1}{{\tt [arXiv:#1]}}}

\providecommand{\url}[1]{\href{#1}{#1}}

}

\providecommand{\adsurl}[1]{} 
\newcommand\SSS{Sect.~}

\nice{ \newcommand\mycaptionfont{\protect\footnotesize} }

\newcommand\gtapprox{\,\lower.6ex\hbox{$\buildrel >\over \sim$} \, }
\newcommand\ltapprox{\,\lower.6ex\hbox{$\buildrel <\over \sim$} \, }
\newcommand\propapprox{\,\lower.6ex\hbox{$\buildrel \propto\over \sim$} \, }

\newcommand\arcs{\ifmmode {'' }\else $'' $\fi}     
\newcommand\arcm{\ifmmode {' }\else $' $\fi}       
\newcommand\ddeg{\ifmmode^\circ\else$^\circ$\fi}    

\newcommand\frtoday{Le\space\number\day\space\ifcase\month\or
  janvier\or f\'evrier\or mars\or avril\or mai\or juin\or
  juillet\or ao\^ut\or septembre\or octobre\or novembre\or 
d\'ecembre\fi\space \number\year}

\newcommand\cqg{ClassQuantGra}   %
\newcommand\hGpc{\mbox{$h^{-1}$ Gpc}}

\newcommand\rC{R_{\mathrm{C}}}  

\newcommand\Omm{\Omega_{\mathrm{m}}}
\newcommand\Omrad{\Omega_{\mathrm{r}}}
\newcommand\Omtot{\Omega_{\mathrm{tot}}} 
\newcommand\OmLam{\Omega_{\Lambda}} 
\newcommand\Omk{\Omega_{\mathrm{k}}} 
\newcommand\rhocrit{\rho_{\mathrm{crit}}}  

\newcommand\notea{^\mathrm{a}}
\newcommand\noteb{^\mathrm{b}}
\newcommand\notec{^\mathrm{c}}
\newcommand\noted{^\mathrm{d}}
\newcommand\notee{^\mathrm{e}}
\newcommand\notef{^\mathrm{f}}
\newcommand\noteg{^\mathrm{g}}
\newcommand\noteh{^\mathrm{h}}

\title{Dark energy as a spatial continuity condition}

\author{Boudewijn F. Roukema\inst{1}
\and Vincent Blanl{\oe}il\inst{2}
}

\institute{Toru\'n Centre for Astronomy, Nicolaus Copernicus University,
ul. Gagarina 11, 87-100 Toru\'n, Poland 
\and IRMA, D\'epartement de Math\'ematiques, Universit\'e de Strasbourg,
7 rue Ren\'e Descartes, 67084 Strasbourg cedex, France
}

\date{\frtoday}

\titlerunning{$\Omega_\Lambda$ as a spatial continuity condition}
\authorrunning{Roukema \& Blanl{\oe}il}

\begin{document}


\newcommand\Nchainsmain{16}
\newcommand\Npergroup{four}

\abstract
{Observational evidence of dark energy that makes
  the Universe nearly flat at the present epoch is
  very strong.}
{ We study the link between spatial continuity and
  dark energy.  }
{ We assume that comoving space is a compact
  3-manifold of constant curvature, described by a
  homogeneous
  Friedman-Lema\^{\i}tre-Robertson-Walker
  metric. We assume that spatial continuity cannot
  be violated, i.e. that the global topology of
  the comoving section of the Universe cannot
  change during post-quantum epochs. }
{ We find that if the Universe was flat and
  compact during early epochs, then the presently
  low values of the radiation and matter densities
  imply that dark energy was created as a spatial
  continuity effect.  Moreover, {\em if the
    Universe is compact, then $\Omtot=1$ is
    dynamically stable}, where $\Omtot$ is the
  total density parameter in units of the critical
  density. }
{ Dark energy was observationally detected as a
  geometrical phenomenon.  It is difficult to
  imagine a simpler explanation for dark energy
  than spatial continuity, finiteness and
  homogeneity. }


\maketitle

\dodouble{ \clearpage } 


\newcommand\tcoeff{
\begin{table}
\caption{\mycaptionfont 
Statistical characteristics of coefficients $a_i$ 
of the dominant ($i$-th order) term 
in the radial and orthogonal components of the residual acceleration $\ddot{r}$ in 
perfectly regular well-proportioned spaces, for approximately isotropic 
displacements $\mathbf{r}$.$\notea$
\label{t-coeff}}
$$\begin{array}{c c c r r r r} \hline  \hline
\mbox{space} \rule[-1.5ex]{0ex}{4.5ex}
& \mbox{term}\noteb
& \parallel/\perp\notec  
& \left<a_i\right>
& \sigma_{\left<a_i\right>}\noted
& \sigma_i\notee
& \gamma_i\notef
\\ \hline 
\rule{0ex}{2.5ex}    
\mbox{3-torus} & (r/L_a)^3 & \parallel &       0.00 &   0.00  & 6.11 &       0.58  \\ 
\mbox{3-torus} & (r/L_a)^3 & \perp &       6.39 & 0.00 & 6.82 & -0.42  \\ 
\mathrm{octahedral} & (r/\rC)^3 & \parallel &        0.01 &       0.01 &       3.18 &      -0.58  \\ 
\mathrm{octahedral} & (r/\rC)^3 & \perp &        3.33 &       0.00 &       1.26 &      -0.42  \\ 
\mathrm{tr. cube} & (r/\rC)^3 & \parallel &       -0.05 &       0.05 &      14.24 &       0.58  \\ 
\mathrm{tr. cube} & (r/\rC)^3 & \perp &       14.94 &       0.02 &       5.63 &      -0.42  \\ 
\mathrm{dodec/num}\noteg & (r/\rC)^5 & \parallel &       -0.30 &       0.53 &     288.29 &       0.75  \\ 
\mathrm{dodec/alg}\noteh & (r/\rC)^5 & \parallel &     0.00 &       0.01 &     288.26 &       0.74 \\ 
\mathrm{dodec/num}\noteg & (r/\rC)^5 & \perp &      286.33 &       0.22 &     121.65 &      -0.37  \\ 

\hline
\end{array}$$
\\
$\notea$ coefficients $a_i$ as defined in 
Eqs~(\protect\ref{e-define-a3-flat}),
(\protect\ref{e-define-a3-S3}), and
(\protect\ref{e-define-a5-S3}); these are approximately constant with respect to $r$;
the constant factor of $Gm/L_a^2$ for the $T^3$ model 
or $Gm/\rC^2$ for the other models has been ignored here; all values
shown are dimensionless \\
$\noteb$ dominant $i$-th power of displacement, as derived in this paper \\
$\notec$ radial $\parallel$ or orthogonal $\perp$ component \\
$\noted$ standard error in the mean
$\sigma_{\left<a_i\right>} =
\sigma_i/\sqrt{N-1}$ for $N \gg 1 $ test particles \\
$\notee$ sample standard deviation \\
$\notef$ sample skewness $\gamma_i = 
      \left<[(a_i- \left<a_i\right>)/\sigma_i]^3 \right>$ \\
$\noteg$ from 70-bit significand numerical calculations using 
      Eq.~(\protect\ref{e-residgrav-S3-Gamma}) \\
$\noteh$ using the algebraic expression in 
      Eq.~(\protect\ref{e-residgrav-Poincare-exact})
\end{table}
}  

\newcommand\fomegathree{
\begin{figure}
\centering 
\includegraphics[width=8cm]{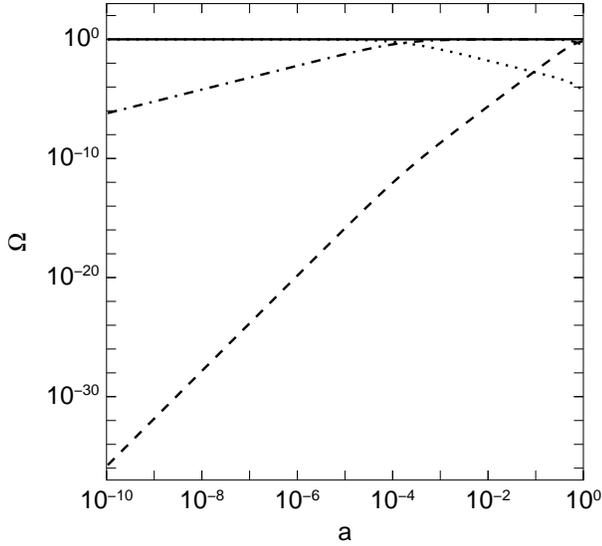}
\caption[]{ \mycaptionfont
Density parameters relative to the critical density from
well into the radiation-dominated epoch at $a=10^{-10}$ 
to the present, showing $\Omtot$ (solid line), 
$\Omrad$ (dotted line), 
$\Omm$ (dot-dashed line), 
$\OmLam$ (dashed line), 
where
${\Omrad}_0 = 4.95 \times 10^{-5}$,
${\Omm}_0 = 0.3$, ${\OmLam}_0 = 0.715$, $w=-1$. See 
Eqs~(\protect\ref{e-omall-evolution}) and
(\protect\ref{e-fried}).
}
\label{f-omega-3}
\end{figure} 
} 

\newcommand\fhthreezthree{
\begin{figure}
\centering 
\includegraphics[width=7cm]{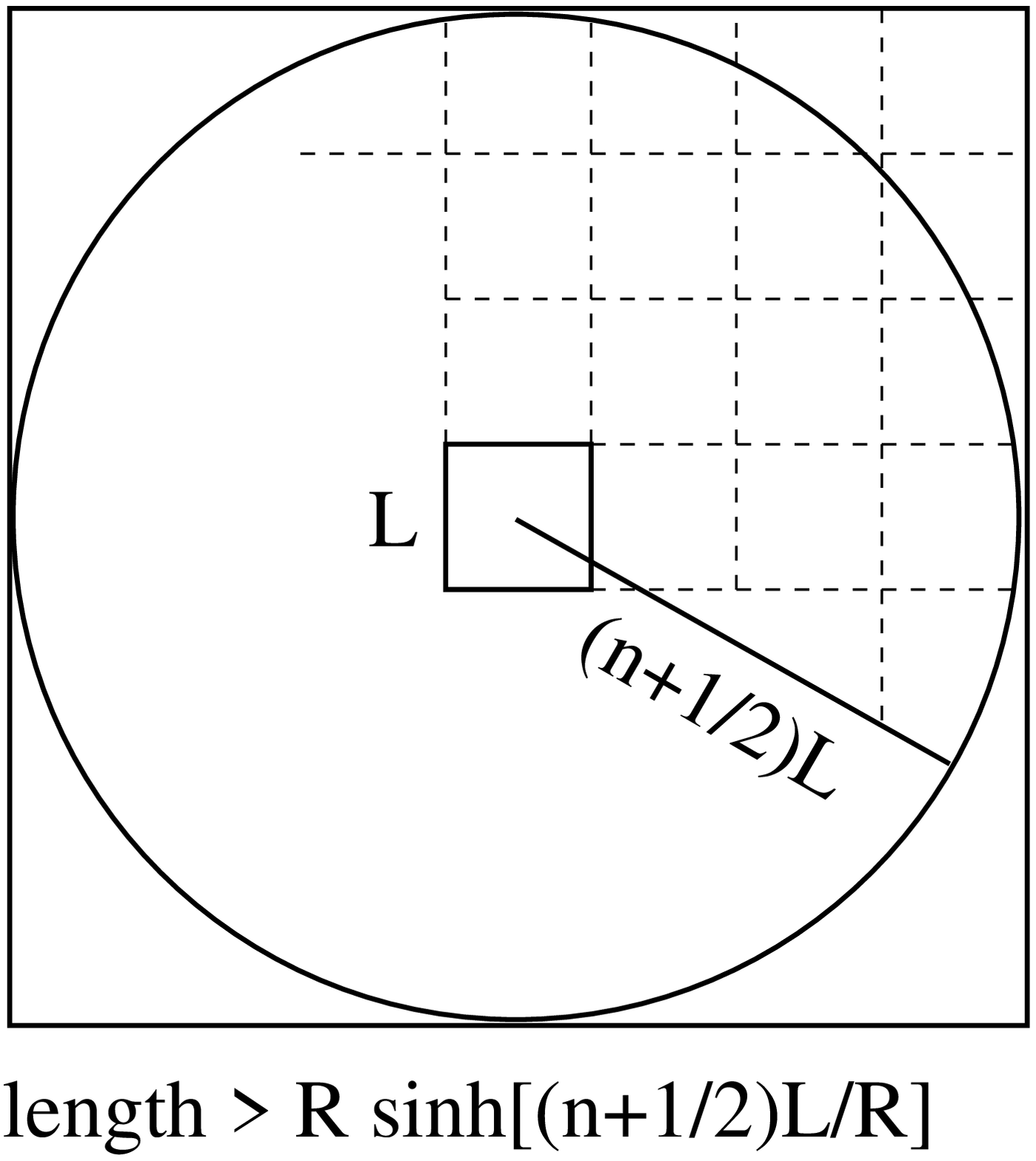}
\caption[]{ \mycaptionfont
\protect\prerefereechanges{Diagram indicating the inconsistency of the would-be
compact, constant curvature 3-manifold
$\mathbb{H}^3/\mathbb{Z}^3$. 
One copy of the fundamental 
domain (FD), at the centre, has a surface area $6A^2$,
implying Eq.~(\protect\ref{e-surfacearea-fd}) for the
surface area of the ``big'' cube composed of 
$(2n+1)^3$ copies of the FD in the universal cover 
(apparent space). However, negative curvature gives
a lower bound to the side lengths of the big cube 
or 
the areas of the surfaces of the big cube, given 
in Eq.~(\protect\ref{e-surfacearea-h3}), which leads
to a contradiction for sufficiently large $n$.}
}
\label{f-h3z3}
\end{figure} 
} 

\newcommand\ffluid{
\begin{figure}
\centering 
\includegraphics[width=7cm]{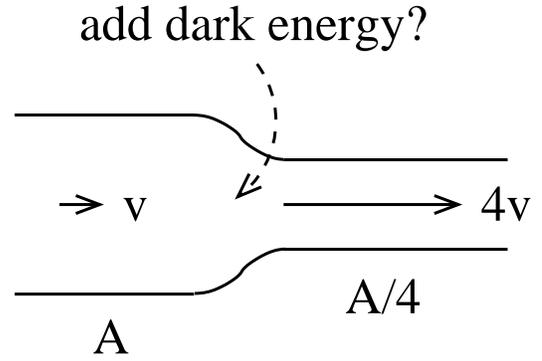}
\caption[]{ \mycaptionfont \protect\prerefereechanges{Schematic
    diagram showing how geometry can induce a physical effect with no
    requirement for ``new physics''. A fluid of constant density flows
    at constant speed $v$ through a pipe of initially constant
    cross-sectional area $A$ through to a bottleneck of new constant
    cross-sectional area $A/4$. Its new speed is $4v$. The change
    in speed can be attributed either to the onset of a dark energy
    epoch, or to geometrical change together with the assumptions of
    spatial continuity, finiteness and homogeneity.}
}
\label{f-fluid}
\end{figure} 
} 

\newcommand\fstability{
\begin{figure}
\centering 
\includegraphics[width=7cm]{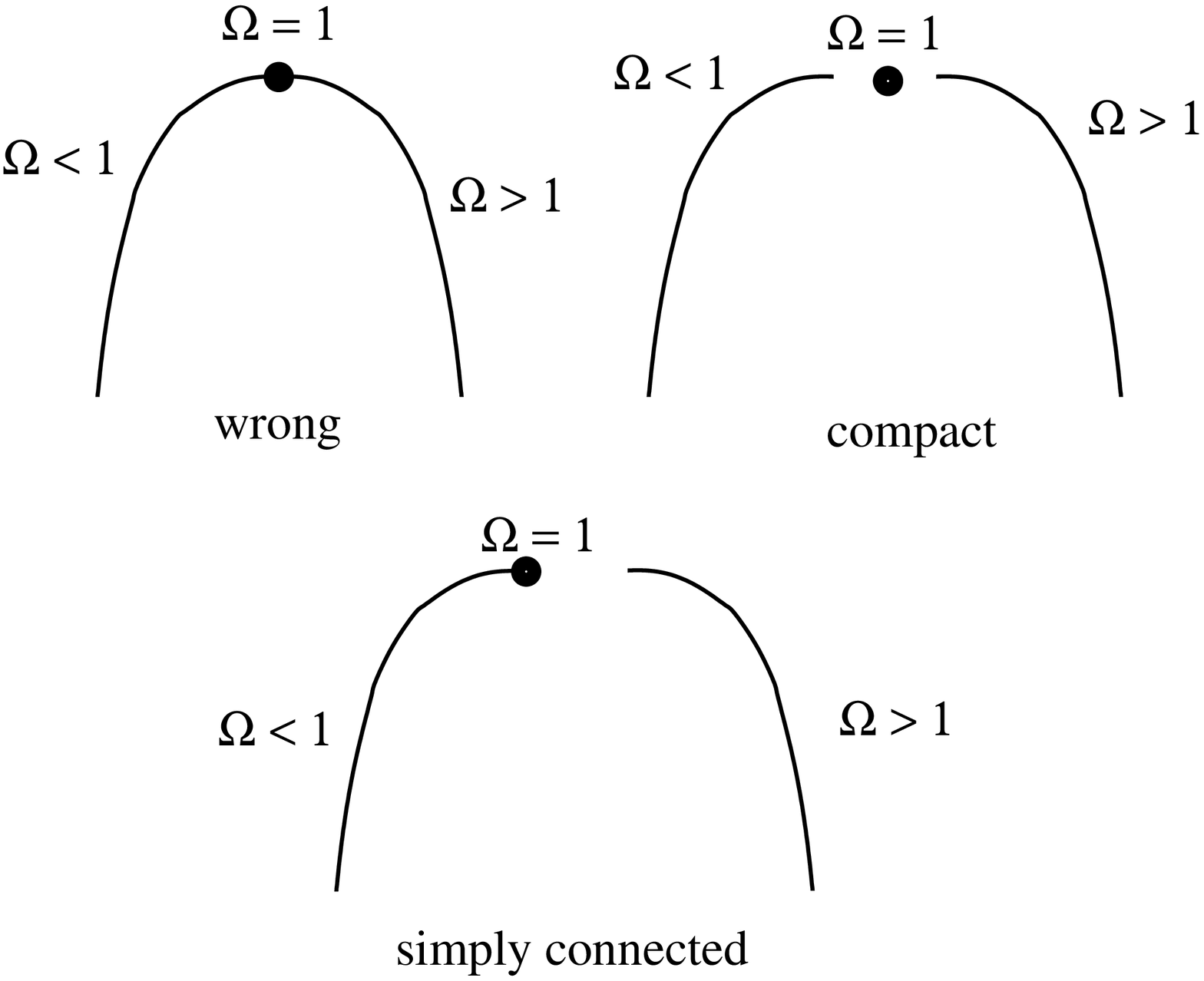}
\caption[]{ \mycaptionfont Schematic
    diagram showing the stability of the total density to
    critical density ratio $\Omtot$ (written $\Omega$ for simplicity) as a
    function of choices of comoving 3-manifolds. Top-left: the usual
    assumption of $\Omtot$ being an unstable point on a continuum
    through all three signs of the curvature is wrong for all choices
    of constant curvature 3-manifolds. Top-right: in the compact case,
    no continuous transformation between different curvatures is possible.
    Bottom: continuous evolution between
    $\mathbb{H}^3$ and $\mathbb{R}^3$ is possible, but not from either
    of these to $S^3$.
}
\label{f-stability}
\end{figure} 
} 


\section{Introduction}  \label{s-intro}

The flatness problem \nocite{Liddle99}(e.g. Sect.~3.1,  {Liddle} 1999,  and references
  therein) has long been a challenge for modern cosmology.
Given the baryon density provided by galaxies as a lower limit to the
total matter-energy density $\Omtot$ 
[expressed in units of the critical density 
$\rhocrit \equiv 3 H^2/(8\pi G)$, where 
$H$ is the Hubble parameter and $G$ is the gravitational constant], 
why is the Universe within an order of
magnitude or two of the critical density, which is the density at
which the Universe is flat?  Inflationary scenarios provide one
possible answer to both this problem and to some other cosmological
problems.  However, during the 1990's, observational evidence
using many different observational strategies
\nocite{FYTY90,FortMD97,ChY97,SCP9812}({Fukugita} {et~al.} 1990; {Fort} {et~al.} 1997; {Chiba} \& {Yoshii} 1997; {Perlmutter} {et~al.} 1999)
established the existence of ``dark energy'', which together
with matter density makes the
Universe nearly flat (total density within about ten percent of the
critical density) at the present epoch. More recent evidence constrains 
the total density to within about one percent of the critical density
\nocite{WMAPSpergel06,ACBAR08}({Spergel} {et~al.} 2007; {Reichardt} {et~al.} 2008). This 
\prerefereechanges{is} the new flatness problem.

Inflationary scenarios may well provide the correct explanation of why
the density of the Universe is presently within a few orders of
magnitude of the critical density. However, for primordial inflation
to lead to a present-day inflationary epoch that takes over from
matter domination just as the matter density is dropping below the
critical density constitutes a new fine-tuning problem.
Many hypotheses have been raised in order to provide an explanation for
dark energy \nocite{CaldwellK09}(e.g.,  {Caldwell} \& {Kamionkowski} 2009)
but the simplest possible explanation seems to have been overlooked.

The Friedman-Lema\^{\i}tre-Robertson-Walker models of the Universe
implicitly assume that the comoving spatial section of the Universe
is a 3-manifold. This assumption does not constitute ``new physics''.
It could better be described as ``implicit physics''.
Does this assumption have any consequences for dark energy?

In \SSS\ref{s-methods}, our assumptions are presented. In
\SSS\ref{s-results}, the consequences are examined.  Discussion and
conclusions are presented in \SSS\ref{s-disc} and \SSS\ref{s-conclu}.

\fomegathree

\section{Method} \label{s-methods}

\subsection{The dark energy/flatness problem} \label{s-meth-DEflatness-prob}

First consider the local properties of
Friedman-Lema\^{\i}tre-Robertson-Walker (FLRW) models.  Here,
``local'' refers to the fact that the Einstein equations describe the
properties in the limit towards any arbitrary point $p$ in comoving
space (averaged on sufficiently ``large'' scales), i.e. properties that
\prerefereechanges{apply at every $p$ in $M$,}
where the comoving space is a constant
curvature Riemannian 3-manifold $M$. Let us write the Friedman
equation as
\begin{eqnarray}
  H^2 &=&  H_0^2 \left[
    {\Omrad}_0 a^{-4} + {\Omm}_0 a^{-3} + 
    {\Omk}_0 a^{-2}   +  {\OmLam}_0 a^{-3-3w}\right]
\label{e-fried}
\end{eqnarray}
where the present values of matter-energy densities 
are written relative to the critical
density $\rhocrit$ with the subscript ``0''
\prerefereechanges{to indicate the present epoch 
$[a(t_0) \equiv a_0 \equiv 1]$},
i.e. the radiation energy density is 
${\Omrad}_0$,
the non-relativistic matter density is 
${\Omm}_0$, 
dark energy is ${\OmLam}_0$,
and the curvature ``density parameter'' 
${\Omk}_0$ is the present day value of curvature 
density parameter $\Omk$ 
defined 
\begin{equation}
\Omk(a) \equiv 1 - \Omrad(a) - \Omm(a) - \OmLam(a)
\end{equation}
at an arbitrary scale factor $a$ 
\nocite{Peebles1993}(cf. Eqs~(13.3), (13.4),  {Peebles} 1993).
We can write the evolution of these components in terms of their present-epoch
values, allowing $w \equiv p/\rho$ to parametrise
a more general dark energy model than that of a cosmological constant:
\begin{eqnarray}
\Omrad &=& \frac{{\Omrad}_0 a^{-4}}{(H/H_0)^2} \nonumber \\
\Omm &=& \frac{{\Omm}_0 a^{-3}}{(H/H_0)^2} \nonumber \\
\prerefereechanges{\Omk}
&=& \frac{( 1 - {\Omm}_0 - {\Omrad}_0 - {\OmLam}_0 ) a^{-2}}{(H/H_0)^2}
\nonumber \\
\OmLam &=& \frac{{\OmLam}_0 a^{-3-3w}}{(H/H_0)^2}.
\label{e-omall-evolution}
\end{eqnarray}
The curvature radius (real for positive curvature, imaginary for
negative curvature) can be written
\begin{equation}
R_C^2 = \left(\frac{c}{H}\right)^2 \frac{1}{(-\Omk)}.
\end{equation}

Figure~\ref{f-omega-3} shows the evolution of the radiation density,
the matter density and dark energy with $w=-1$, i.e. a cosmological
constant, from well before the Universe became matter-dominated to the
present. The cosmological constant at $a=10^{-10}$ had to be $\OmLam
\approx 1.82 \times 10^{-37}$ at that epoch in order to have just the
right value in order to make the Universe close to flat
\prerefereechanges{today}. 
The matter density at that epoch was $\Omm(10^{-10}) \approx 6.42 \times
10^{-7}$. 
How was it possible for the physical content of the Universe at $a=10^{-10}$
to ``know'' that dark energy, which at the time was 30 orders of magnitude
less dense than matter, should be within a few percent of the required
values in order that it would start dominating the Universe just when 
matter density would drop significantly below unity?

This is the new fine-tuning problem, which is not solved by
inflationary scenarios. The specific numbers cited here can be
modified slightly by including a neutrino density or other estimates
of ${\Omrad}_0$, ${\Omm}_0$ and/or ${\OmLam}_0$ consistent with
present observational constraints, but the fine-tuning problem is only
altered slightly.  Inflationary scenarios are normally thought to
finish much earlier than $a = 10^{-10}$, so the required
``conspiracy'' between $\Omm$ and $\OmLam$ is even worse than what is
shown in Fig.~\ref{f-omega-3}.  For example, if we ignore the
matter-dominated epoch in order to make an order-of-magnitude
calculation and write $a \sim (t/t_0)^{1/2}$, then $(H/H_0) \sim
(t/t_0)^{-1}$ and a post-inflationary epoch of $t \sim 10^{-30}$~s
gives $\OmLam \sim {\OmLam}_0 (H/H_0)^2$ $\sim {\OmLam}_0 (t/t_0)^{-2}$
$\sim 10^{-95}.$ How was it possible that a dark energy
density nearly 100 orders of magnitude smaller than the critical density
was ``waiting'' in preparation for a future epoch when it would take
over from matter in order to flatten the Universe at just the right moment?

\subsection{The assumption that comoving space is a 3-manifold}

The FLRW models of the Universe
implicitly assume that the comoving spatial section of the Universe is
a 3-manifold. An additional assumption that is often made implicitly,
without any discussion nor theoretical or observational evidence, is
that this 3-manifold is topologically trivial (has a 
\prerefereechanges{trivial} $\pi_1$
homotopy group). However, this assumption is arbitrary. The covering
space of comoving space for an FLRW model is either $\mathbb{H}^3$,
$\mathbb{R}^3$, or ${S}^3$ for negative, zero, or positive curvature
respectively. The different possible 
fundamental groups of 
\prerefereechanges{holonomy} transformations $\Gamma$ for these
three covering spaces give the FLRW constant curvature 
Riemannian 3-manifolds of interest in
cosmology, i.e. $\mathbb{H}^3/\Gamma$,
$\mathbb{R}^3/\Gamma$, or ${S}^3/\Gamma$ respectively for the three
signs of the curvature. The groups $\Gamma$ are, in general, 
\prerefereechanges{non-isomorphic for different} curvatures.

The arbitrary nature of this trivial topology assumption is
first known to have been raised by Karl Schwarzschild \nocite{Schw00,Schw98}({Schwarzschild} 1900, 1998).
It has been analysed most intensively during the last decade and 
a half \nocite{LaLu95,Lum98,Stark98,LR99,BR99}(see  {Lachi\`eze-Rey} \& {Luminet} 1995; {Luminet} 1998; {Starkman} 1998; {Luminet} \& {Roukema} 1999; {Blanl{\oe}il} \& {Roukema} 2000,  and references therein)
and different observational strategies for measuring cosmic topology
(the topology of the 3-manifold of comoving space) have been developed
and classified \nocite{ULL99b,LR99,Rouk02topclass,RG04}(e.g.,  {Uzan} {et~al.} 1999; {Luminet} \& {Roukema} 1999; {Roukema} 2002; {Rebou\c{c}as} \& {Gomero} 2004).

Some theoretical work that might constitute the basis for deciding
which 3-manifold should be favoured by a theory of quantum cosmology
has been carried out \nocite{Masafumi96,CarlipSurya04}({Masafumi} 1996; {Anderson} {et~al.} 2004). 
A recent heuristic result is that of the dynamical
effect of cosmic topology in the presence of density perturbations.
A residual weak limit gravitational effect
in the presence of a density perturbation
selects well-proportioned spaces in general, and the Poincar\'e
dodecahedral space ${S}^3/I^*$ in particular, to be special in the sense
of being ``better balanced''
\nocite{RBBSJ06,RR09}({Roukema} {et~al.} 2007; {Roukema} \& {R\'o\.za\'nski} 2009). Recent empirical analyses have mostly focussed
on the Wilkinson Microwave Anisotropy Probe (WMAP) all-sky maps of the
cosmic microwave background. The results are presently inconclusive,
ranging from excluding detectable cosmic topology
\nocite{CSSK03,KeyCSS06}({Cornish} {et~al.} 2004; {Key} {et~al.} 2007), preferring a simply connected infinite flat
space \nocite{NJ07,LR08}({Niarchou} \& {Jaffe} 2007; {Lew} \& {Roukema} 2008), preferring the Poincar\'e dodecahedral space
over simply connected infinite flat space
\nocite{LumNat03,Aurich2005a,Aurich2005b,Gundermann2005,Caillerie07,RBSG08,RBG08}({Luminet} {et~al.} 2003; {Aurich} {et~al.} 2005a, 2005b; {Gundermann} 2005; {Caillerie} {et~al.} 2007; {Roukema} {et~al.} 2008a, 2008b),
or preferring the regular, flat torus $T^3$ over simply connected
infinite flat space
\nocite{WMAPSpergel,Aurich07align,Aurich08a,Aurich08b,Aurich09a}({Spergel} {et~al.} 2003; {Aurich} {et~al.} 2007; {Aurich} 2008; {Aurich} {et~al.} 2008, 2009).

It is clear that even if cosmic microwave background maps from Planck
Surveyor give evidence against both the Poincar\'e space and 3-torus
models, there is at present no strong theoretical nor empirical argument
requiring comoving space to be either simply connected nor multiply
connected.

\subsection{The assumption of spatial continuity}

Now consider the assumption that {\em the global topology of the
  comoving section of the Universe cannot change during post-quantum
  epochs.} Could an infinite $\mathbb{R}^3$ comoving space suddenly
become a finite $T^3$? Or vice versa?  Or could space evolve from
one finite space to another? 
{
For example, would the following be physically
reasonable during an epoch well past the quantum epoch? 
Changes 
corresponding to ``cutting''
vast square
(comoving) gigaparsec surfaces of the Universe and then ``sticking'' them
back together in a new way, in order to transform comoving space from a
flat space whose fundamental domain is (for example) the hexagonal prism (before
cutting) to $T^3$ (after ``sticking'' back together the cut surfaces),
would have had to have occurred.
}
Or could a $T^3$ comoving space be cut apart, stretched and curved
slightly and then stuck back together to make one of the spherical spaces
${S}^3/\Gamma$, where $\Gamma$ is a group of holonomy transformations of
${S}^3$? At the quantum epoch, a superposition of various 3-manifold
states whose probabilities depend on operators might be reasonable,
but not at post-quantum epochs.  

In principle, during post-quantum epochs for a
comoving observer, black holes modify the topology of comoving
space. However, astrophysically realistic black holes are relatively
tiny and isolated from one another from the point of view of
``average'' comoving space 
\prerefereechanges{
on scales of hundreds of megaparsecs and above}.
Moreover, they form mostly at very recent 
epochs. Structure formation in the standard
cosmological model does not give any hint that black holes could lead
to modification of the topology of the comoving spatial section of the
Universe.
\prerefereechanges{Therefore, for the purposes of this work, let us cut out
a neighbourhood of radius 100 Schwarzschild radii around every
black hole (whether stellar, intermediate or supermassive) and
replace it by an homogeneous solid ball.
}

Hence, let us make the physically reasonable assumption that global
topology change at post-quantum epochs is physically impossible. In other
words, we assume comoving spatial continuity. 

\subsection{Compact comoving space}

Objects of infinite size or mass are generally disliked in physics.
Comoving space itself might be an exception to the rule. However,
finite, continuous space without any boundaries provides a more
conventional physical model than infinite space \nocite{LevSS98a}({Levin} {et~al.} 1998). 
For this reason,
we focus on compact (finite) 3-manifolds
{(except for \SSS\ref{s-stability})}. These include
multiply connected flat and hyperbolic 3-manifolds, and spherical
3-manifolds of any topology.

\subsection{Intuitive tools for thinking about 3-manifolds}

See \nocite{LaLu95,Lum98,Stark98,LR99,BR99}{Lachi\`eze-Rey} \& {Luminet} (1995); {Luminet} (1998); {Starkman} (1998); {Luminet} \& {Roukema} (1999); {Blanl{\oe}il} \& {Roukema} (2000) for general reviews about
3-manifolds of interest to cosmology. Here, we note in particular that
there are at least three different ways of thinking about a 
constant curvature 2-manifold, of which two are relatively easy to
use for 3-manifolds. We focus on the two dimensional flat torus $T^2$
as an illustrative example.
\begin{list}{(\roman{enumi})}{\usecounter{enumi}}
\item $T^2$ can be thought of embedded in $\mathbb{R}^3$, i.e. as 
``the surface of a doughnut with a hole in the middle'', but given
its own intrinsic (flat) metric rather than the metric of $\mathbb{R}^3$.
The continuity of $T^2$ is obvious, but the constant curvature of its
metric is less obvious.
\item $T^2$ can be thought of as a fundamental domain, i.e. a square
(if it is a regular $T^2$ model) with identified edges. The constant 
(zero) curvature is now obvious, but continuity may seem less obvious.
\item $T^2$ can be thought of as the covering space or apparent space
$\mathbb{R}^2$, which contains (infinitely) many copies of $T^2$. 
The continuity and constant zero curvature are now obvious, at the
cost of the existence of multiple images of any single physical point
of space.
\end{list}
For 3-manifolds, method (i) would require intuition
\prerefereechanges{
in at least $\mathbb{R}^4$ (and at most $\mathbb{R}^7$; 
\nocite{Whitney36}{Whitney} 1936). 
 For understanding the physical consequences of
comoving space being a 3-manifold, it is generally useful to switch
between methods (ii), thinking of a 3-manifold as a fundamental domain
(polyhedron) with faces identified in a certain way, and (iii) thinking
of the covering space ($\mathbb{H}^3$, $\mathbb{R}^3$ or $S^3$) tiled
by identically shaped (in a metric sense) copies of the fundamental
domain. The 3-manifold itself is, of course, physically identical
no matter which method is used for modelling it cognitively.}

\section{Results} \label{s-results}

\subsection{Zero curvature during post-quantum epochs}
\label{s-res-zero}

\subsubsection{$T^3$}

First consider the flat case. In particular, let us start
with $T^3 = \mathbb{R}^3/\mathbb{Z}^3$. That is, let us suppose
that at some post-quantum epoch, comoving space was
$T^3$. For example, this was probably determined by the end of
the quantum epoch.
Suppose also that there is no present-epoch dark energy, 
i.e. $\OmLam = 0$.

Now suppose that at much more recent epochs, the various components
of matter-energy density dropped in density so that the density
came to be dominated by the matter density alone, i.e. we arrive
at the observed value 
${\Omm}_0 \approx 0.3$. By the Friedman equation, this implies
that the curvature became significantly negative. Hence, our assumption
of no topology change implies that comoving space evolved from 
$T^3 = \mathbb{R}^3/\mathbb{Z}^3$ to 
$\mathbb{H}^3/\mathbb{Z}^3$.

However, $\mathbb{H}^3/\mathbb{Z}^3$ as a constant curvature 3-manifold
does not exist 
\nocite{Best1971}(see e.g.  {Best} 1971).
\prerefereechanges{An elementary} proof is given in Appendix~\ref{s-T3-H3}. 
Either comoving space became
highly inhomogeneous on the largest scales possible, or comoving space
remained homogeneous and remained $\mathbb{R}^3/\mathbb{Z}^3$. Since
we assume an FLRW model, we ignore the former possibility and are forced
to accept the latter. 

How is it possible for space to have remained
homogeneous $\mathbb{R}^3/\mathbb{Z}^3$ even though the matter density
dropped to ${\Omm}_0 \approx 0.3$?
The obvious explanation is the constant of integration allowed by
solving the Einstein equations. {\em A cosmological constant, or more
generally, dark energy $\OmLam$, allows the equations to find a solution that
conserves spatial continuity and homogeneity even though the evolution
of matter-energy density alone in the absence of dark energy would
require a violation of either homogeneity or spatial continuity.}
In other words, if comoving space is finite, continuous and close
to homogeneous and retains these three conditions as it evolves, 
then dark energy is a stretching effect necessarily 
induced by these conditions.
Hereafter, we refer to this as ``continuity dark energy''.

This is somewhat analogous to solving a general physical equation
with a given set of boundary conditions. Multiply connected space
is sometimes described as having ``periodic boundary conditions''.
However, this is a somewhat misleading method of thinking about the nature of 
multiply connected space,
since there is no physical repetition of space-time events, only
an effect similar to mirroring. 
Rather than referring to ``continuity dark energy'' as a ``boundary
condition effect'', it would be more
physically useful to refer to it as a ``spatial continuity effect''.

\fstability

This can be rewritten as follows. If the
Universe was $T^3$ when it exited the quantum epoch, then since that
epoch, it has at different epochs been in one of the following three
states:
\begin{list}{(\roman{enumi})}{\usecounter{enumi}}
\item highly inhomogeneous\footnote{For example, thinking of 
the fundamental domain, keep the ``six faces'' 
unchanged and only stretch the ``interior'' of the fundamental
domain.}
\item homogeneous
and ``supported'' as flat space by one or more forms of ``bottom-up'' 
matter-energy density (e.g. photons, or dark matter particles and baryons, 
or more exotic particles or fields at earlier epochs); or 
\item homogeneous and flattened by 
``top-down'' dark energy which compensated exactly for
``bottom-up'' forms of matter-energy density that were insufficient
to ``support'' flatness.
\end{list}

Since ${S}^3/\mathbb{Z}^3$ does not exist either (the fundamental groups
of spherical constant curvature 3-manifolds are finite, but $\mathbb{Z}$ 
is infinite), this reasoning applies
even if some exotic form of matter-energy density would (in the absence
of dark energy) 
\prerefereechanges{try to}
increase the density to super-critical during certain
epochs. Similar arguments apply 
for other flat compact spaces than $T^3$ \nocite{LaLu95}(e.g., Sect.~6.2 {Lachi\`eze-Rey} \& {Luminet} 1995).


{A useful way to think of this is
in terms of the (comoving) fundamental domain. Provided we
have the assumptions of homogeneity, spatial continuity and
compactness, the curvature can change from one negative value
to another or one positive value to another, but the sign
(negative, zero, or positive) cannot change. 
To change the sign
of curvature, at least one of these assumptions would have to
be violated.}

\subsubsection{Stability of $\Omtot=1$} \label{s-stability}

It is often stated \nocite{Liddle99}(e.g. Sect.~3.1,  {Liddle} 1999)  that 
$\Omtot=1$ is an unstable point for FLRW models. This statement is
partially true if the topology of comoving space is trivial. If
comoving space is $\mathbb{H}^3$, then it can smoothly evolve to 
$\mathbb{R}^3$, or vice versa, without violating spatial continuity
nor requiring large inhomogeneities to be created. Given the Friedman
equation and the standard behaviour expected of 
matter-density components, it is not easy to change from a negatively
curved space to a perfectly flat space or vice-versa. However, 
spatial continuity does not prevent this. Exotic fields could in principle
allow this evolution without violating either homogeneity or continuity.

On the other hand, evolving from $\mathbb{R}^3$ to ${S}^3$ would
require infinite comoving space to be suddenly wrapped up into a
finite hypersphere using an extremely inhomogeneous transformation
(radially inhomogeneous with respect to a given ``centre''), with the
addition of a new point at the antipode, or the placing of a black
hole there.  The opposite transformation would be equally unappealing.
In other words, even with the assumption of trivial topology, 
the continuity requirement
prevents evolution between $\Omtot \le 1$ and
$\Omtot > 1$. In this sense, $\Omtot$ might be described as
semi-stable or asymmetrically stable.
 
Now consider compact flat space. For simplicity, consider
$T^3$. 
Now suppose that, for example, 
density perturbations shift the average total density to slightly 
higher or lower than the critical density at a given cosmological time.
This would {\em not} cause the Universe to shift further and further away
from $\Omtot=1$, since continuity and homogeneity would force dark energy
to (on average) compensate for the slight (temporary?) curvature. 
Perturbations or
changes in the dominant matter-energy components of the Universe cannot
change $T^3 = \mathbb{R}^3/\mathbb{Z}^3$ into either
$\mathbb{H}^3/\mathbb{Z}^3$ or
${S}^3/\mathbb{Z}^3$, since neither of 
\prerefereechanges{the latter two}
spaces exist. 
Similar arguments apply for the other compact, flat spaces.

Hence, if the Universe is compact, then {\em $\Omtot=1$ is dynamically 
stable.} Compact comoving flat space cannot evolve away from $\Omtot=1$
without violating either homogeneity or 
spatial continuity. The disagreement between this result and previous
statements in the literature is that previous work did not consider
the fundamentally distinct nature of 3-manifolds of different curvatures.
The stability of $\Omtot$ for the different cases 
is shown schematically in Fig.~\ref{f-stability}.

This link between cosmic topology and dynamics should not be confused
with the residual gravity effect \nocite{RBBSJ06,RR09}({Roukema} {et~al.} 2007; {Roukema} \& {R\'o\.za\'nski} 2009). These constitute 
two very different mechanisms linking topology and dynamics.

\subsection{Positive curvature during post-quantum epochs}
\label{s-res-pos}

Now consider the positively curved case, where at some post-quantum
epoch, comoving space was $S^3/\Gamma$. The choice of $\Gamma$ is
unlikely to play a role for the ``creation'' of dark energy, so let us
adopt the Poincar\'e dodecahedral space $S^3/I^*$ as suggested
empirically for the present epoch by the WMAP data, and theoretically
by the residual gravity effect.  Is it possible for this
model to evolve by the present epoch to the present observed value
${\Omm}_0 \approx 0.3$ in the absence of dark energy, i.e.  with
${\OmLam} = 0$? In the absence of adding a further exotic dark energy
component, the Friedman equation already prevents evolution from
positive curvature to negative curvature. 

However, if some exotic component were expected to ``push'' space from
positive to negative curvature, then this would imply that comoving
space evolved from $S^3/I^*$ to $\mathbb{R}^3/I^*$ to
$\mathbb{H}^3/I^*$ as the curvature changed. Our assumption of no
topology change prevents this: the binary icosahedral group $I^*$ is a
finite group, but the fundamental group $\Gamma$ of
$\mathbb{R}^3/\Gamma$ or $\mathbb{H}^3/\Gamma$ must be infinite.
Clearly, the constant curvature 3-manifolds $\mathbb{R}^3/I^*$ and
$\mathbb{H}^3/I^*$ do not exist.

Nevertheless, the addition of an exotic dark energy component that in
turn requires continuity dark energy would constitute a more
complicated model rather than a simpler model. The present dark energy
would be a continuity effect but a new primordial dark energy component
would be required. At least for the simplest models of the Universe,
dark energy does not arise as a consequence of continuity in a positively
curved model.

\subsection{Negative curvature during post-quantum epochs}
\label{s-res-neg}

One optimal member of the set of 3-manifolds 
is the Weeks space \nocite{WeeksPhD85,Fagundes93}({Weeks} 1985; {Fagundes} 1993), which minimises the volume
of negatively curved constant curvature 3-manifolds for a fixed
curvature radius.
Suppose that the quantum epoch selected the Weeks space or another
hyperbolic compact 3-manifold. 
Since the present estimate of the matter-energy density {\em in the absence
of dark energy} is ${\Omtot}_0 \approx {\Omm}_0 \approx 0.3$, the 
evolution from primordial epochs to the present would {\em not} 
require any violation of homogeneity nor spatial continuity, since
the evolution would be from $\mathbb{H}^3/\Gamma$ 
\prerefereechanges{during a primordial epoch to
$\mathbb{H}^3/\Gamma$ at the present}
for the same group of holonomy transformations 
$\Gamma$, with only an evolution in the curvature radius.

Similarly to the positive curvature case, in order for spatial
continuity to induce dark energy if space is $\mathbb{H}^3/\Gamma$, an
additional component of the matter-energy density that would tend to
force the total density from below critical to above critical would
have to exist. In other words, for ``continuity dark energy'' to exist
in a negatively curved space, an additional, as yet unknown form of
matter-energy density would also need to exist.

\ffluid

Again, this would constitute a more complicated model rather than a simpler
model. By Occam's razor, compactness, homogeneity, spatial continuity
{and the observed dark energy}
favour a flat, compact model 
in order for dark energy to arise as a spatial
continuity effect without requiring the addition of any 
``new physics''.

\section{Discussion} \label{s-disc}

\prerefereestart
\subsection{A physical analogy: fluid flow of an homogeneous, incompressible
material}

Figure~\ref{f-fluid} shows a loose analogy for dark energy as a spatial
continuity condition. The flow of a fluid composed of 
an homogeneous incompressible material from a length of pipe with a wide
cross-section to a length of pipe with a narrow cross-section necessarily
causes the speed of the fluid to increase in proportion to the change
in cross-sectional area. The apparent creation of ``dark energy'' causing
the fluid to accelerate is purely a continuity effect. No new model of
particle physics nor braneworld model is required to explain the acceleration
of fluid in a pipe that narrows.

This analogy is not perfect. Instead of comoving space changing its
shape, it retains its shape, while the matter density drops below what
is required by the Einstein equations in order to conserve
the sign of the curvature. Nevertheless, the analogy may help
illustrate why spatial continuity provides a simple physical
understanding of what we refer to as dark energy.

\prerefereestop

\subsection{Exact flatness versus approximate flatness}

The surface of the Earth is not an exact 2-sphere, and the orbits
of the planets in our Solar System are not exact circles. 
It is equally unrealistic for the Universe to have had a curvature
exactly equal to zero everywhere, in the mathematical sense of exactness.
Density perturbations certainly exist today, and at small enough
scales, particles do not consist of 
\prerefereechanges{
a genuinely uniform fluid.}

On the other hand, it is quite possible that topological evolution
during the quantum epoch led to the 3-manifold of comoving space being
$T^3$. The residual gravity effect selects the Poincar\'e
dodecahedral space ${S}^3/I^*$ as a better balanced space
than $T^3$ \nocite{RR09}({Roukema} \& {R\'o\.za\'nski} 2009), but for the sake of argument,
let us suppose that the unknown theory
of the quantum epoch led to comoving space being $T^3$.
In this case, how do we reconcile the nature of $T^3$ as an idealised
mathematical object being perfectly flat and a more realistic
physical model which is very nearly flat, but contains 
some tiny perturbations and most likely has an ``average'' 
curvature (depending on the way that the average is calculated)
which is ``slightly'' different from zero? In fact, there is no
contradiction. From a topological point of view, $T^3$ with
density perturbations is a mathematically valid 3-manifold. 
What happens when we apply the Friedman equation? 
To the best of our experimental knowledge,
the Einstein equations
provide an excellent approximation to reality. However, interpreting
the Friedman equation assuming perfect homogeneity and a ``slightly''
non-zero curvature would imply that the curvature evolves further
and further away from zero. Yet, as noted above, 
this is not possible without violating
either homogeneity or spatial continuity. Retaining spatial 
continuity and approximate homogeneity 
solves the dilemma: a tiny amount of stretching with the properties
of dark energy
would result in order to keep the average curvature
close to zero.

So the difference between idealised perfect flatness and approximate
flatness is only a semantic problem that occurs if curvature is
discussed without reference to topology.

If comoving space is $T^3$ and close to 
homogeneous as assumed in the FLRW models, then in this sense 
it ``is flat'' and will always stay ``flat'' unless a global topology
change occurs. However, the particular values of the average
curvature at a given epoch may differ from zero and 
will depend on the averaging method and on the details of how
information on the spatial continuity requirement is spread
throughout comoving space, i.e. on the details of how 
continuity dark energy responds to the changes in matter-energy
density. So the same space could be considered ``flat in a topological
sense'' while being ``non-flat'' in terms of $\Omtot$.

On the other hand, if comoving space ``is approximately flat'' in the
sense that it is a compact, non-zero curvature 3-manifold,
i.e. ${S}^3/\Gamma$ or $\mathbb{H}^3/\Gamma$ with a small average
curvature at a given epoch, then this is physically very different
from being ``topologically flat''.  No exotic form of matter-energy
density, including an inflaton, can make it become ``topologically
flat'' or switch it to a 3-manifold of the opposite curvature, unless
we drop the assumption of spatial continuity.

\subsection{The present-epoch fine-tuning problem (coincidence problem)} 
\label{e-disc-new-fine-tuning}

Continuity dark energy does not require any conspiracy.
If we consider the $T^3$ model, then 
the ``knowledge'' encoded at $a= 10^{-10}$ was the ``knowledge''
that comoving space is $T^3$ and that any average deviations from
perfect flatness ``needed'' to be compensated. The precise way
in which this occurs (e.g. a field theory) is unknown, but it must occur
unless we wish to drop approximate homogeneity or spatial continuity.
The real evolution of $\OmLam$ with $a$ might be quite different,
but it would have to evolve in a way that space remains ``approximately''
flat. The physical way in which ``approximately'' is encoded in
comoving space should determine the way in which $\OmLam$ really
evolved with $a$.

\subsection{Possible consequences for galaxy formation}

The Friedman equation together with spatial continuity do not
constrain in what way continuity dark energy information is
communicated throughout comoving space.  They only require that this
occurs. Presumably, the information is to some degree represented
everywhere locally, and changes in the information are
transmitted at (or below) the speed of the space-time conversion
constant $c$. 

Did the stretching process occur uniformly?
Continuity dark energy could reasonably have
occurred inhomogeneously on large-scale-structure scales,
i.e. at about 100 comoving megaparsecs
and below, provided that the (global) continuity of comoving space
and approximate homogeneity conditions were satisfied.
As density perturbations collapsed into 
the cosmic web, especially filaments and massive clusters at the knots
where the filaments intersect, it would seem reasonable that
the stretching occurred more in the voids and less along filaments
and at knots, since higher density regions are more tightly bound
than lower density regions.
This may provide the solution to the 
``void phenomenon'' \nocite{Peebles01Voids}({Peebles} 2001), without requiring
any modification of gravity.
Other puzzles in galaxy formation \nocite{PeriPuzzles09}(e.g.  {Perivolaropoulos} 2008)
for the concordance model \nocite{CosConcord95}({Ostriker} \& {Steinhardt} 1995) could also potentially
be resolved by continuity dark energy and/or offer
clues for distinguishing continuity dark energy from the 
exotic dark energy models required in a simply connected,
infinite flat space or in curved spaces.

\subsection{Consequences for primordial inflationary scenarios}

For a flat, compact universe, since dark energy as a continuity effect
explains why the Universe is exactly flat in the 3-manifold sense, it
also explains why the Universe is ``close'' to flat today, i.e. it
resolves the order-of-magnitude flatness problem.  However, the other
arguments in favour of primordial inflation scenarios (the horizon
problem, the magnetic monopoles problem) remain unchanged, so
primordial inflation is certainly not excluded.  

Moreover, in a flat compact space, the continuity argument also
applies during primordial (post-quantum) epochs.  If the dominant
matter-energy density at an early epoch entered a phase where in the
absence of dark energy the curvature would have become negative, then space
must have been stretched unless continuity or homogeneity were
violated.  The question of the nature of the inflaton then becomes a
search for the dominant matter-energy components {\em prior} to the
primordial inflationary epoch, since the inflaton itself would just be 
a spatial continuity effect.  It is also interesting that
\nocite{Linde04topo}{Linde} (2004) has argued that under certain conditions, a
compact hyperbolic or flat universe would be preferred in inflationary
scenarios, i.e. in agreement with the spatial continuity requirement
in the case of flat models, though not in the case of non-flat models.

\section{Conclusion} \label{s-conclu}
 The key assumptions required here are that 
\begin{list}{(\roman{enumi})}{\usecounter{enumi}}
\item comoving space was a compact 3-manifold after exiting the quantum
epoch;
\item during post-quantum epochs, the Universe cannot
  have been ripped apart and pasted back together again in a different
  way (spatial continuity); and
\item the Universe has remained close to homogeneous at all post-quantum
epochs.
\end{list}
Comoving spatial continuity provides an explanation for the
present epoch cosmological constant without requiring any
``new physics'', provided that the 3-manifold of comoving
space is a compact, flat 3-manifold, such as $T^3$  
[see {\SSS}6.2, \nocite{LaLu95}{Lachi\`eze-Rey} \& {Luminet} (1995), for other compact, flat 3-manifolds].
The same effect would shift the
search for explaining a primordial inflationary epoch from
\prerefereechanges{the search for}
 an ``inflaton'' to 
\prerefereechanges{the search for a}
pre-inflationary
matter-energy density that in the absence of the continuity
requirement would cause the Universe to become hyperbolic.

This effect does not require multiple connectedness to be observable
today by direct methods.  However, the missing fluctuations problem in COsmic
Background Explorer (COBE) and WMAP maps of the cosmic microwave
background has become stronger with the release of the five-year WMAP
data \nocite{Copi07,Copi09}(e.g.  {Copi} {et~al.} 2007, 2008) and would be simply explained by
living in an observably compact space.  Several analyses of the WMAP
data favour a Poincar\'e dodecahedral space ($S^3/I^*$) model over the
infinite, simply connected flat model
\nocite{LumNat03,Aurich2005a,Aurich2005b,Gundermann2005,Caillerie07,RBSG08,RBG08}({Luminet} {et~al.} 2003; {Aurich} {et~al.} 2005a, 2005b; {Gundermann} 2005; {Caillerie} {et~al.} 2007; {Roukema} {et~al.} 2008a, 2008b). On the other hand,
recent work has shown that a $T^3$ model
\nocite{WMAPSpergel,Aurich07align,Aurich08a,Aurich08b,Aurich09a}({Spergel} {et~al.} 2003; {Aurich} {et~al.} 2007; {Aurich} 2008; {Aurich} {et~al.} 2008, 2009) also
provides a better fit to the empirical constraints than the infinite, simply
connected flat model. Given that dark energy is explained as a continuity
effect in a compact flat space without requiring any ``new physics'',
Occam's razor suggests that a $T^3$ model of comoving side length
about 12{\hGpc} \nocite{Aurich09a}(e.g.  {Aurich} {et~al.} 2009) and fundamental directions
in the directions listed in Table~1 of \nocite{Aurich08a}{Aurich} (2008) may be
our present best model of comoving space.

\begin{acknowledgements}
Some of this work was carried out within the framework of the European
Associated Laboratory ``Astrophysics Poland-France''.
Use was made 
of the
Centre de Donn\'ees astronomiques de Strasbourg 
(\url{http://cdsads.u-strasbg.fr}),
the GNU {\sc Octave} command-line, high-level numerical computation software 
(\url{http://www.gnu.org/software/octave}),
and the GNU {\sc plotutils} plotting package.
BFR thanks the Observatoire de Strasbourg for hospitality during
which part of this work was carried out.

%
%
%

\end{acknowledgements}

\subm{ \clearpage }

\nice{
%

}


\fhthreezthree

\appendix
\section{Does the constant curvature 3-manifold $\mathbb{H}^3/\mathbb{Z}^3$ 
exist?}
\label{s-T3-H3} 

Suppose that the constant curvature 3-manifold
$\mathbb{H}^3/\mathbb{Z}^3$ exists. By symmetry, let us
represent it using $\mathbb{R}^3$ as a coordinate system for
the covering space $\mathbb{H}^3$, with an appropriate metric
of curvature radius $R$.  Let the size of the fundamental
domain be $L$.  We then have $\mathbb{R}^3$ tiled by an
infinite grid of copies of the fundamental domain, which is a
regular cube of volume $L^3$ at the origin of our coordinate
system.

Consider the cube in $\mathbb{R}^3$ composed of $(2n+1)^3$
copies of the fundamental domain, centred at the origin,
as shown in Fig.~\ref{f-h3z3}. The
surface area of this cube consists of $6(2n+1)^2$ faces of
the fundamental domain. Since every copy of the fundamental
domain has exactly the same shape, in the metric sense, this
surface area is
\begin{equation}
A_{\mathrm{big}} = 6(2n+1)^2 A,
\label{e-surfacearea-fd}
\end{equation}
where
$A$ is the surface area of one face of 
the fundamental domain of $\mathbb{H}^3/\mathbb{Z}^3$.
\prerefereechanges{Since space is hyperbolic, 
$A> L^2$ and $L^2/A$ are fixed values, independent of
$n$.}

However, every one of the $6(2n+1)$ faces of copies of the fundamental domain
on the surface of this large cube is at a distance of 
\prerefereechanges{at least $(n+1/2)L$ from} 
the origin. Hence, the surface area of the large cube must
be 
\begin{eqnarray}
A_{\mathrm{big}}' &>& 4 \pi  
\prerefereechanges{    \{ R\; \mathrm{sinh}[(n+1/2)L/R] \}^2. }
\label{e-surfacearea-h3}
\end{eqnarray}
\prerefereechanges{Since $A, L$ and $R$ are fixed values, a large value of $N$
can be found such that }
\begin{equation}
n > N \Rightarrow
A_{\mathrm{big}}'(n) > A_{\mathrm{big}}(n).
\end{equation}
However, $A_{\mathrm{big}}' = A_{\mathrm{big}}$ by definition.
Hence, $\mathbb{H}^3/\mathbb{Z}^3$ cannot exist.


\begin{thebibliography}{}

\bibitem[{Anderson}, {Carlip}, {Ratcliffe}, {Surya},  \& {Tschantz} 2004]{CarlipSurya04}
{Anderson}, M., {Carlip}, S., {Ratcliffe}, J.~G., {Surya}, S., \& {Tschantz},  S.~T. 2004, \cqg, 21, 729, \eprint{gr-qc/0310002}

\bibitem[{Aurich} 2008]{Aurich08a}
{Aurich}, R. 2008, \cqg, 25, 225017, \eprint{0803.2130}

\bibitem[{Aurich}, {Janzer}, {Lustig}, \&  {Steiner} 2008]{Aurich08b}
{Aurich}, R., {Janzer}, H.~S., {Lustig}, S., \& {Steiner}, F. 2008, Classical  and Quantum Gravity, 25, 125006, \eprint{0708.1420}

\bibitem[{Aurich}, {Lustig}, \&  {Steiner} 2005a]{Aurich2005a}
{Aurich}, R., {Lustig}, S., \& {Steiner}, F. 2005a, \cqg, 22,  3443, \eprint{astro-ph/0504656}

\bibitem[{Aurich}, {Lustig}, \&  {Steiner} 2005b]{Aurich2005b}
{Aurich}, R., {Lustig}, S., \& {Steiner}, F. 2005b, \cqg, 22,  2061, \eprint{astro-ph/0412569}

\bibitem[{Aurich}, {Lustig}, \& {Steiner} 2009]{Aurich09a}
{Aurich}, R., {Lustig}, S., \& {Steiner}, F. 2009, ArXiv e-prints,  \eprint{0903.3133}

\bibitem[{Aurich}, {Lustig}, {Steiner}, \&  {Then} 2007]{Aurich07align}
{Aurich}, R., {Lustig}, S., {Steiner}, F., \& {Then}, H. 2007, \cqg, 24, 1879,  \eprint{astro-ph/0612308}

\bibitem[{Best} 1971]{Best1971}
{Best}, L.~A. 1971, Canad.J.Math., 23, 451

\bibitem[{Blanl{\oe}il} \& {Roukema} 2000]{BR99}
{Blanl{\oe}il}, V., \& {Roukema}, B.~F., eds. 2000, ``Cosmological Topology in  Paris 1998'' (Paris: Blanl{\oe}il \& Roukema), \eprint{astro-ph/0010170}

\bibitem[{Caillerie}, {Lachi{\`e}ze-Rey}, {Luminet},  {Lehoucq}, {Riazuelo}, \& {Weeks} 2007]{Caillerie07}
{Caillerie}, S., {Lachi{\`e}ze-Rey}, M., {Luminet}, J.~., {et al.} 2007, \aap, 476, 691, \eprint{0705.0217v2}

\bibitem[{Caldwell} \& {Kamionkowski} 2009]{CaldwellK09}
{Caldwell}, R.~R., \& {Kamionkowski}, M. 2009, ArXiv e-prints,  \eprint{0903.0866}

\bibitem[{Chiba} \& {Yoshii} 1997]{ChY97}
{Chiba}, M., \& {Yoshii}, Y. 1997, \apj, 489, 485

\bibitem[{Copi}, {Huterer}, {Schwarz}, \&  {Starkman} 2007]{Copi07}
{Copi}, C.~J., {Huterer}, D., {Schwarz}, D.~J., \& {Starkman}, G.~D. 2007,  \prd, 75, 023507, \eprint{astro-ph/0605135}

\bibitem[{Copi}, {Huterer}, {Schwarz}, \&  {Starkman} 2008]{Copi09}
{Copi}, C.~J., {Huterer}, D., {Schwarz}, D.~J., \& {Starkman}, G.~D. 2008,  ArXiv e-prints, \eprint{0808.3767}

\bibitem[{Cornish}, {Spergel}, {Starkman}, \&  {Komatsu} 2004]{CSSK03}
{Cornish}, N.~J., {Spergel}, D.~N., {Starkman}, G.~D., \& {Komatsu}, E. 2004,  Phys. Rev. Lett., 92, 201302, \eprint{astro-ph/0310233}

\bibitem[{Fagundes} 1993]{Fagundes93}
{Fagundes}, H.~V. 1993, Physical Review Letters, 70, 1579

\bibitem[{Fort}, {Mellier}, \& {Dantel-Fort} 1997]{FortMD97}
{Fort}, B., {Mellier}, Y., \& {Dantel-Fort}, M. 1997, \aap, 321, 353

\bibitem[{Fukugita}, {Yamashita}, {Takahara}, \&  {Yoshii} 1990]{FYTY90}
{Fukugita}, M., {Yamashita}, K., {Takahara}, F., \& {Yoshii}, Y. 1990, \apjl,  361, L1

\bibitem[{Gundermann} 2005]{Gundermann2005}
{Gundermann}, J. 2005, arXiv preprints, \eprint{astro-ph/0503014}

\bibitem[{Key}, {Cornish}, {Spergel}, \&  {Starkman} 2007]{KeyCSS06}
{Key}, J.~S., {Cornish}, N.~J., {Spergel}, D.~N., \& {Starkman}, G.~D. 2007,  \prd, 75, 084034, \eprint{astro-ph/0604616}

\bibitem[{Lachi\`eze-Rey} \& {Luminet} 1995]{LaLu95}
{Lachi\`eze-Rey}, M., \& {Luminet}, J. 1995, \physrep, 254, 135,  \eprint{gr-qc/9605010}

\bibitem[{Levin}, {Scannapieco}, \& {Silk} 1998]{LevSS98a}
{Levin}, J., {Scannapieco}, E., \& {Silk}, J. 1998, \prd, 58, 103516,  \eprint{astro-ph/9802021}

\bibitem[{Lew} \& {Roukema} 2008]{LR08}
{Lew}, B., \& {Roukema}, B.~F. 2008, \aap, 482, 747, \eprint{0801.1358}

\bibitem[{Liddle} 1999]{Liddle99}
{Liddle}, A.~R. 1999, in High Energy Physics and Cosmology, 1998 Summer School,  ed. A.~{Masiero}, G.~{Senjanovic}, \& A.~{Smirnov}, 260--+

\bibitem[{Linde} 2004]{Linde04topo}
{Linde}, A. 2004, Journal of Cosmology and Astro-Particle Physics, 10, 4,  \eprint{arXiv:hep-th/0408164}

\bibitem[{Luminet} \& {Roukema} 1999]{LR99}
{Luminet}, J., \& {Roukema}, B.~F. 1999, in NATO ASIC Proc. 541: Theoretical  and Observational Cosmology. Publisher: Dordrecht: Kluwer,, 117,  \eprint{astro-ph/9901364}

\bibitem[{Luminet}, {Weeks}, {Riazuelo}, {Lehoucq}, \&  {Uzan} 2003]{LumNat03}
{Luminet}, J., {Weeks}, J.~R., {Riazuelo}, A., {Lehoucq}, R., \& {Uzan}, J.  2003, \nat, 425, 593, \eprint{astro-ph/0310253}

\bibitem[{Luminet} 1998]{Lum98}
{Luminet}, J.-P. 1998, Acta Cosmologica, XXIV-1, 105, \eprint{gr-qc/9804006}

\bibitem[{Masafumi} 1996]{Masafumi96}
{Masafumi}, S. 1996, \prd, 53, 6902, \eprint{gr-qc/9603002v1}

\bibitem[{Niarchou} \& {Jaffe} 2007]{NJ07}
{Niarchou}, A., \& {Jaffe}, A. 2007, Physical Review Letters, 99, 081302,  \eprint{astro-ph/0702436}

\bibitem[{Ostriker} \& {Steinhardt} 1995]{CosConcord95}
{Ostriker}, J.~P., \& {Steinhardt}, P.~J. 1995, ArXiv Astrophysics e-prints,  \eprint{astro-ph/9505066}

\bibitem[{Peebles} 1993]{Peebles1993}
{Peebles}, P.~J.~E. 1993, {Principles of physical cosmology} (Princeton Series  in Physics, Princeton, NJ: Princeton University Press)

\bibitem[{Peebles} 2001]{Peebles01Voids}
{Peebles}, P.~J.~E. 2001, \apj, 557, 495, \eprint{arXiv:astro-ph/0101127}

\bibitem[{Perivolaropoulos} 2008]{PeriPuzzles09}
{Perivolaropoulos}, L. 2008, ArXiv e-prints, \eprint{0811.4684}

\bibitem[{Perlmutter}, {Aldering}, {Goldhaber},  {Knop}, {Nugent}, {Castro}, {Deustua}, {Fabbro}, {Goobar}, {Groom}, {Hook},  {Kim}, {Kim}, {Lee}, {Nunes}, {Pain}, {Pennypacker}, {Quimby}, {Lidman},  {Ellis}, {Irwin}, {McMahon}, {Ruiz-Lapuente}, {Walton}, {Schaefer}, {Boyle},  {Filippenko}, {Matheson}, {Fruchter}, {Panagia}, {Newberg}, {Couch}, \& {The  Supernova Cosmology Project} 1999]{SCP9812}
{Perlmutter}, S., {Aldering}, G., {Goldhaber}, G., {et al.} 1999, \apj, 517, 565, \eprint{astro-ph/9812133}

\bibitem[{Rebou\c{c}as} \& {Gomero} 2004]{RG04}
{Rebou\c{c}as}, M.~J., \& {Gomero}, G.~I. 2004, Braz. J. Phys., 34, 1358,  \eprint{astro-ph/0402324}

\bibitem[{Reichardt}, {Ade}, {Bock}, {Bond},  {Brevik}, {Contaldi}, {Daub}, {Dempsey}, {Goldstein}, \& {et al.} 2008]{ACBAR08}
{Reichardt}, C.~L., {Ade}, P.~A.~R., {Bock}, J.~J., {et al.} 2008, ArXiv e-prints, 801, \eprint{0801.1491}

\bibitem[{Roukema} 2002]{Rouk02topclass}
{Roukema}, B.~F. 2002, in {Marcel Grossmann IX Conference on General  Relativity}, eds V.G. Gurzadyan, R.T. Jantzen and R. Ruffini, Singapore:  World Scientific,, p. 1937, \eprint{astro-ph/0010189}

\bibitem[{Roukema}, {Bajtlik}, {Biesiada},  {Szaniewska}, \& {Jurkiewicz} 2007]{RBBSJ06}
{Roukema}, B.~F., {Bajtlik}, S., {Biesiada}, M., {Szaniewska}, A., \&  {Jurkiewicz}, H. 2007, \aap, 463, 861, \eprint{astro-ph/0602159}

\bibitem[{Roukema}, {Buli\'nski}, \&  {Gaudin} 2008b]{RBG08}
{Roukema}, B.~F., {Buli\'nski}, Z., \& {Gaudin}, N.~E. 2008b,  \aap, 492, 673, \eprint{0807.4260}

\bibitem[{Roukema}, {Buli\'nski},  {Szaniewska}, \& {Gaudin} 2008a]{RBSG08}
{Roukema}, B.~F., {Buli\'nski}, Z., {Szaniewska}, A., \& {Gaudin}, N.~E.  2008a, \aap, 486, 55, \eprint{0801.0006}

\bibitem[{Roukema} \& {R\'o\.za\'nski} 2009]{RR09}
{Roukema}, B.~F., \& {R\'o\.za\'nski}, P.~T. 2009, \aap, in press,  \eprint{0902.3402}

\bibitem[{Schwarzschild} 1900]{Schw00}
{Schwarzschild}, K. 1900, Vier.d.Astr.Gess, 35, 337

\bibitem[{Schwarzschild} 1998]{Schw98}
{Schwarzschild}, K. 1998, \cqg, 15, 2539

\bibitem[{Spergel}, {Bean}, {Dor\'e}, {Nolta},  {Bennett}, {Hinshaw}, {Jarosik}, {Komatsu}, {Page}, {Peiris}, {Verde},  {Barnes}, {Halpern}, {Hill}, {Kogut}, \& {et al.} 2007]{WMAPSpergel06}
{Spergel}, D.~N., {Bean}, R., {Dor\'e}, O., {et al.} 2007, \apjs, 170, 377, \eprint{astro-ph/0603449}

\bibitem[{Spergel}, {Verde}, {Peiris}, {Komatsu},  {Nolta}, {Bennett}, {Halpern}, {Hinshaw}, {Jarosik}, {Kogut}, {Limon},  {Meyer}, {Page}, {Tucker}, {Weiland}, {Wollack}, \& {Wright} 2003]{WMAPSpergel}
{Spergel}, D.~N., {Verde}, L., {Peiris}, H.~V., {et al.} 2003, \apjs, 148, 175,  \eprint{astro-ph/0302209}

\bibitem[{Starkman} 1998]{Stark98}
{Starkman}, G.~D. 1998, \cqg, 15, 2529

\bibitem[{Uzan}, {Lehoucq}, \& {Luminet} 1999]{ULL99b}
{Uzan}, J.-P., {Lehoucq}, R., \& {Luminet}, J.-P. 1999, in {Proc. of the  XIX$^{\rm th}$ Texas meeting, Paris 14--18 December 1998, Eds. E. Aubourg, T.  Montmerle, J. Paul and P. Peter, article n$^{\rm o}$ 04/25, Nuclear Physics B  (Proc. Suppl.) No. 80}, \eprint{gr-qc/0005128}

\bibitem[{Weeks} 1985]{WeeksPhD85}
{Weeks}, J. 1985, PhD thesis, Princeton University (1985)

\bibitem[{Whitney} 1936]{Whitney36}
{Whitney}, H. 1936, The Annals of Mathematics, Second Series, 37, 645

\end{thebibliography}
\end{document}